\documentstyle[aps,prl,preprint,floats,epsfig]{revtex}

\providecommand{\BDD}{\mbox{$B^0\to D^{(*)+}D^{(*)-}$}}
\providecommand{\BDSTDST}{\mbox{$B^0\to D^{*+}D^{*-}$}}
\providecommand{\BBARDSTDST}{\mbox{$\bar{B}^0\to D^{*+}D^{*-}$}}
\providecommand{\BDDST}{\mbox{$B^0\to D^{*\pm}D^{\mp}$}} 
\providecommand{\BOR}{\mbox{$B^0\to D^{(*)+}D^{-}$}}
\providecommand{\BDPDM}{\mbox{$B^0\to D^{+}D^{-}$}}
\providecommand{\DPDMtyp}{\mbox{$\DP\to\KPIPI,\,\DM\to\CKPIPI$}}
\providecommand{\DDSTtyp}{\mbox{$\DST\to\DZ\pi^+,\,\DZ\to\KPIPIZERO,\,\DM\to\CKPIPI$}}
\providecommand{\KS}{\mbox{$K^0_{\rm S}$}}
\providecommand{\PSIKS}{\mbox{$J/\psi\KS$}}
\providecommand{\BPSIKS}{\mbox{$B^0\to \PSIKS$}}
\providecommand{\BR}{\mbox{${\cal B}$}}
\providecommand{\Br}{\mbox{${\cal B}$}}
\providecommand{\BB}{\mbox{$B\bar{B}$}}
\providecommand{\EPEM}{\mbox{$e^+e^-$}}
\providecommand{\CC}{\mbox{$c\bar{c}$}}
\providecommand{\QQ}{\mbox{$q\bar{q}$}}
\providecommand{\UPSFOUR}{\mbox{$\Upsilon(4{\rm S})$}}
\providecommand{\UFOURS}{\mbox{$\Upsilon(4{\rm S})$}}
\providecommand{\cleoii}{\mbox{CLEO II}}
\providecommand{\cleoiiv}{\mbox{CLEO II.V}}
\providecommand{\BDSD}{\mbox{$B^0\to D^{(*)+}_sD^{(*)-}$}}
\providecommand{\DST}{\mbox{$D^{*+}$}}
\providecommand{\DSTP}{\mbox{$D^{*+}$}}
\providecommand{\DSTM}{\mbox{$D^{*-}$}}
\providecommand{\DZ}{\mbox{$D^{0}$}}
\providecommand{\DP}{\mbox{$D^{+}$}}
\providecommand{\DM}{\mbox{$D^{-}$}}
\providecommand{\DOPP}{\mbox{$D^{(*)+}$}}
\providecommand{\DOPM}{\mbox{$D^{(*)-}$}}
\providecommand{\BZ}{\mbox{$B^{0}$}}
\providecommand{\BZBAR}{\mbox{$\bar{B}^{0}$}}
\providecommand{\SLOWPI}{\mbox{$\pi_{\rm s}$}}
\providecommand{\KPI} {\mbox{${K^-}\pi^+ $}}
\providecommand{\KPIPIZERO} {\mbox{${K^-}\pi^+\pi^0 $}}
\providecommand{\KPIPIPI} {\mbox{${K^-}\pi^+\pi^+\pi^- $}}
\providecommand{\KSHORTPIPI} {\mbox{${K}^0_{\rm S}\pi^+\pi^- $}}
\providecommand{\KSHORTPIPIPIZERO} {\mbox{${K}^0_{\rm S}\pi^+\pi^-\pi^0 $}}

\providecommand{\KPIPI} {\mbox{${K^-}\pi^+\pi^+ $}}
\providecommand{\KSHORTPI} {\mbox{${K}^0_{\rm S}\pi^+ $}}
\providecommand{\KSHORTPIPIZERO} {\mbox{${K}^0_{\rm S}\pi^+\pi^0 $}}
\providecommand{\KPIPIPIZERO} {\mbox{${K^-}\pi^+\pi^+\pi^0 $}}
\providecommand{\KSHORTPIPIPI} {\mbox{${K}^0_{\rm S}\pi^+\pi^+\pi^- $}}
\providecommand{\KKPI} {\mbox{${K^-K^+}\pi^+ $}}
\providecommand{\KKPIPIZERO} {\mbox{${K^-K^+}\pi^+\pi^0 $}}
\providecommand{\CKPIPI} {\mbox{${K^+}\pi^-\pi^- $}}

\providecommand{\CHIM}{\mbox{$\chi^2_{M}$}}
\providecommand{\LSL}{\mbox{$L/\sigma(L)$}}
\providecommand{\DELE}{\mbox{$\Delta E$}}
\providecommand{\MB}{\mbox{$M(B)$}}
\providecommand{\HELA}{\mbox{$\theta_{\rm H}$}}
\providecommand{\COSHEL}{\mbox{$\cos\HELA$}}
\providecommand{\THRA}{\mbox{$\theta_{\rm T}$}}
\providecommand{\COSTHR}{\mbox{$\cos\THRA$}}
\providecommand{\TVA}{\mbox{$\theta_{tr}$}}
\providecommand{\TV}{\mbox{$\cos\TVA$}}
\providecommand{\TTMF}{\mbox{$\times 10^{-4}$}} 
\providecommand{\TTMT}{\mbox{$\times 10^{-3}$}} 
\providecommand{\BRBDSTDST}{\mbox{$9.9{ }^{+4.2}_{-3.3}\pm 1.2$}} 
\providecommand{\BRBDSTDSTss}{\mbox{$9.9{ }^{+4.2}_{-3.3}\ {\rm [stat.]}\pm 1.2\ {\rm [syst.]}$}} 
\providecommand{\BRBDDST}{\mbox{$2.1{ }^{+2.4}_{-1.7}\pm 0.5$}} 
\providecommand{\BRBDPDM}{\mbox{$3.0{ }^{+3.3 +0.8}_{-2.1 -0.6}$}} 
\providecommand{\ULBDSTDST}{\mbox{---}}
\providecommand{\ULBDDST}{\mbox{$6.3$}} 
\providecommand{\ULBDPDM}{\mbox{$9.4$}} 
\providecommand{\BKGDBDSTDST}{\mbox{$0.42\pm0.04\pm0.13$}} 
\providecommand{\BKGDBDDST}{\mbox{$1.68\pm0.07\pm0.24$}} 
\providecommand{\BKGDBDPDM}{\mbox{$0.46\pm0.03\pm0.10$}} 
\providecommand{\EFFBDSTDST}{\mbox{$8.2\pm 2.9$}} %
\providecommand{\EFFBDDST}{\mbox{$11.0\pm 1.7$}} %
\providecommand{\EFFBDPDM}{\mbox{$5.4\pm 1.0$}} 
\providecommand{\ACP}{\mbox{$0.49{}^{+0.41}_{-0.34}\pm 0.02$}}
\providecommand{\ANINETY}{\mbox{$0.11$}}
\providecommand{\SCFBDSTDST}{\mbox{$(7.3\pm 2.2)\TTMT$}}
\providecommand{\SCFBDDSTII}{\mbox{$(4.7\pm 1.2)\TTMT$}}
\providecommand{\SCFBDDSTIIV}{\mbox{$(4.3\pm 0.7)\TTMT$}}
\providecommand{\SCFBDPDM}{\mbox{$(4.0\pm 0.9)\TTMT$}}

\begin{document}


\tighten

\title{Study of the Decays {\mathversion{bold}\BDD} }  

\author{
E.~Lipeles, M.~Schmidtler, A.~Shapiro, W.~M.~Sun,
A.~J.~Weinstein, and F.~W\"{u}rthwein%
\thanks{Permanent address: Massachusetts Institute of Technology, Cambridge, MA 02139.}}
\address{
California Institute of Technology, Pasadena, California 91125}
\author{
D.~E.~Jaffe, G.~Masek, H.~P.~Paar, E.~M.~Potter, S.~Prell,
and V.~Sharma}
\address{
University of California, San Diego, La Jolla, California 92093}
\author{
D.~M.~Asner, A.~Eppich, T.~S.~Hill, R.~J.~Morrison, and H.~N.~Nelson}
\address{
University of California, Santa Barbara, California 93106}
\author{
R.~A.~Briere}
\address{
Carnegie Mellon University, Pittsburgh, Pennsylvania 15213}
\author{
B.~H.~Behrens, W.~T.~Ford, A.~Gritsan, J.~Roy, and J.~G.~Smith}
\address{
University of Colorado, Boulder, Colorado 80309-0390}
\author{
J.~P.~Alexander, R.~Baker, C.~Bebek, B.~E.~Berger, K.~Berkelman,
F.~Blanc, V.~Boisvert, D.~G.~Cassel, M.~Dickson, P.~S.~Drell,
K.~M.~Ecklund, R.~Ehrlich, A.~D.~Foland, P.~Gaidarev, L.~Gibbons,
B.~Gittelman, S.~W.~Gray, D.~L.~Hartill, B.~K.~Heltsley,
P.~I.~Hopman, C.~D.~Jones, D.~L.~Kreinick, M.~Lohner,
A.~Magerkurth, T.~O.~Meyer, N.~B.~Mistry, E.~Nordberg,
J.~R.~Patterson, D.~Peterson, D.~Riley, J.~G.~Thayer,
P.~G.~Thies, B.~Valant-Spaight, and A.~Warburton}
\address{
Cornell University, Ithaca, New York 14853}
\author{
P.~Avery, C.~Prescott, A.~I.~Rubiera, J.~Yelton, and J.~Zheng}
\address{
University of Florida, Gainesville, Florida 32611}
\author{
G.~Brandenburg, A.~Ershov, Y.~S.~Gao, D.~Y.-J.~Kim, and R.~Wilson}
\address{
Harvard University, Cambridge, Massachusetts 02138}
\author{
T.~E.~Browder, Y.~Li, J.~L.~Rodriguez, and H.~Yamamoto}
\address{
University of Hawaii at Manoa, Honolulu, Hawaii 96822}
\author{
T.~Bergfeld, B.~I.~Eisenstein, J.~Ernst, G.~E.~Gladding,
G.~D.~Gollin, R.~M.~Hans, E.~Johnson, I.~Karliner, M.~A.~Marsh,
M.~Palmer, C.~Plager, C.~Sedlack, M.~Selen, J.~J.~Thaler,
and J.~Williams}
\address{
University of Illinois, Urbana-Champaign, Illinois 61801}
\author{
K.~W.~Edwards}
\address{
Carleton University, Ottawa, Ontario, Canada K1S 5B6 \\
and the Institute of Particle Physics, Canada}
\author{
R.~Janicek  and P.~M.~Patel}
\address{
McGill University, Montr\'eal, Qu\'ebec, Canada H3A 2T8 \\
and the Institute of Particle Physics, Canada}
\author{
A.~J.~Sadoff}
\address{
Ithaca College, Ithaca, New York 14850}
\author{
R.~Ammar, A.~Bean, D.~Besson, R.~Davis, N.~Kwak, and X.~Zhao}
\address{
University of Kansas, Lawrence, Kansas 66045}
\author{
S.~Anderson, V.~V.~Frolov, Y.~Kubota, S.~J.~Lee, R.~Mahapatra,
J.~J.~O'Neill, R.~Poling, T.~Riehle, A.~Smith, and J.~Urheim}
\address{
University of Minnesota, Minneapolis, Minnesota 55455}
\author{
S.~Ahmed, M.~S.~Alam, S.~B.~Athar, L.~Jian, L.~Ling,
A.~H.~Mahmood,%
\thanks{Permanent address: University of Texas - Pan American, Edinburg, TX 78539.}
M.~Saleem, S.~Timm, and F.~Wappler}
\address{
State University of New York at Albany, Albany, New York 12222}
\author{
A.~Anastassov, J.~E.~Duboscq, K.~K.~Gan, C.~Gwon, T.~Hart,
K.~Honscheid, D.~Hufnagel, H.~Kagan, R.~Kass, T.~K.~Pedlar,
H.~Schwarthoff, J.~B.~Thayer, E.~von~Toerne, and M.~M.~Zoeller}
\address{
Ohio State University, Columbus, Ohio 43210}
\author{
S.~J.~Richichi, H.~Severini, P.~Skubic, and A.~Undrus}
\address{
University of Oklahoma, Norman, Oklahoma 73019}
\author{
S.~Chen, J.~Fast, J.~W.~Hinson, J.~Lee, N.~Menon, D.~H.~Miller,
E.~I.~Shibata, I.~P.~J.~Shipsey, and V.~Pavlunin}
\address{
Purdue University, West Lafayette, Indiana 47907}
\author{
D.~Cronin-Hennessy, Y.~Kwon,%
\thanks{Permanent address: Yonsei University, Seoul 120-749, Korea.}
A.L.~Lyon, and E.~H.~Thorndike}
\address{
University of Rochester, Rochester, New York 14627}
\author{
C.~P.~Jessop, H.~Marsiske, M.~L.~Perl, V.~Savinov, D.~Ugolini,
and X.~Zhou}
\address{
Stanford Linear Accelerator Center, Stanford University, Stanford,
California 94309}
\author{
T.~E.~Coan, V.~Fadeyev, Y.~Maravin, I.~Narsky, R.~Stroynowski,
J.~Ye, and T.~Wlodek}
\address{
Southern Methodist University, Dallas, Texas 75275}
\author{
M.~Artuso, R.~Ayad, C.~Boulahouache, K.~Bukin, E.~Dambasuren,
S.~Karamov, G.~Majumder, G.~C.~Moneti, R.~Mountain, S.~Schuh,
T.~Skwarnicki, S.~Stone, G.~Viehhauser, J.C.~Wang, A.~Wolf,
and J.~Wu}
\address{
Syracuse University, Syracuse, New York 13244}
\author{
S.~Kopp}
\address{
University of Texas, Austin, TX  78712}
\author{
S.~E.~Csorna, I.~Danko, K.~W.~McLean, Sz.~M\'arka, and Z.~Xu}
\address{
Vanderbilt University, Nashville, Tennessee 37235}
\author{
R.~Godang, K.~Kinoshita,%
\thanks{Permanent address: University of Cincinnati, Cincinnati, OH 45221}
I.~C.~Lai, and S.~Schrenk}
\address{
Virginia Polytechnic Institute and State University,
Blacksburg, Virginia 24061}
\author{
G.~Bonvicini, D.~Cinabro, S.~McGee, L.~P.~Perera, and G.~J.~Zhou}
\address{
Wayne State University, Detroit, Michigan 48202}
 
\author{(CLEO Collaboration)}

\date{\today}
\maketitle

\begin{abstract} 

 The decays \BDSTDST, \BDDST\ and \BDPDM\ are studied 
in $9.7\times 10^6$ $\UPSFOUR\to\BB$ decays accumulated with the 
CLEO detector. 
We determine
$\BR(\BDSTDST) = (\BRBDSTDSTss)\TTMF$ and limit 
$\BR(\BDDST) < \ULBDDST\TTMF$ and
$\BR(\BDPDM) < \ULBDPDM\TTMF$ at 90\% confidence level (CL). We also perform
the first angular analysis of the \BDSTDST\ decay and 
determine that the 
$CP$-even fraction of the final state is greater than 
\ANINETY\ at 90\% CL.
Future measurements of the time dependence of
these decays may be useful for the investigation of $CP$ violation
in neutral $B$ meson decays.

\end{abstract}



\section{Introduction\label{sec:intro}}

 The first observation of $CP$ violation outside
the neutral kaon system~\cite{ref:epsk,ref:epsprime} may well be
a non-zero difference in the rates of 
$B^0\to\PSIKS$ and $\bar{B}^0\to\PSIKS$ decays~\cite{ref:opal.cdf.aleph}.
Such a measurement would be an important test of the
Standard Model mechanism for $CP$ violation as described  by 
the CKM quark-mixing matrix~\cite{ref:CKM}. 
In the Standard Model, the CKM matrix is unitary; for
three quark generations, this property can be represented
as a triangle in the complex plane with internal angles
$\alpha$, $\beta$ and $\gamma$~\cite{ref:pdg}.
Asymmetries in the rate of neutral $B$ meson decays to $CP$ eigenstates 
that occur via the Cabibbo-favored 
$\bar{b}\to \bar{c}W^+; W^+\to c\bar{s}$ ({\it eg.}, \BPSIKS) process
are expected to be proportional to $\sin 2\beta$.
In contrast to the decay \BPSIKS, 
for the Cabibbo-suppressed processes~\footnote{
\BDD\ denotes the decays 
\BDPDM, $\BZ\to\DST\DM$, $\BZ\to\DP\DSTM$ and \BDSTDST.
\BDDST\ denotes the sum of  $\BZ\to\DST\DM$ and  $\BZ\to\DP\DSTM$.
} \BDD, 
the weak phase difference between the tree ($\bar{b}\to\bar{c}c\bar{d}$) 
and penguin ($\bar{b}\to\bar{d}c\bar{c}$) 
amplitudes 
may be appreciable~\cite{ref:ciuchini,ref:sanda.xing}.
In the absence of a strong interaction phase difference between
the tree  and penguin
\BDD\ amplitudes, the magnitude of the asymmetry would be also 
proportional to $\sin 2\beta$.  
The decay rate asymmetry of \BDSTDST\ decays would also be
proportional to $\sin 2\beta$ but may suffer from dilution due 
to the $P$-wave ($CP$-odd) component of the \DSTP\DSTM\ final 
state~\cite{ref:rosner,ref:pham}. 
The relative 
$CP$-even and $CP$-odd components of the \BDSTDST\ decay can be determined
by an angular analysis~\cite{ref:dunietz} that removes any such
dilution.

 Measurements of rate asymmetries in the decays {\BDD}
may provide a means to resolve the four-fold ambiguity in $\beta$ inherent in a 
measurement of  $\sin 2\beta$ from \BPSIKS\ 
decays~\cite{ref:rosner,ref:explain,ref:babar}.
Comparison of the measured asymmetries in \BPSIKS\ and \BDPDM\ decays
may allow partial resolution of the ambiguity in the determination
of $\beta$ if the sign of the ratio of the tree and penguin 
amplitudes of \BDPDM\ decays can be ascertained~\cite{ref:grossman.quinn}.
\BZ\ and \BZBAR\ mesons decay to the same $D^{*+}D^-$ final
state with amplitudes of comparable magnitude and
significant interference between them is 
possible~\cite{ref:xing.plb443,ref:xing.prd61}.
As for \BDPDM, the asymmetry between the rates of
$\BZ\to D^{*+}D^-$ and $\BZBAR\to D^{*-}D^+$ is directly proportional
to $\sin 2\beta$ in the absence of strong phase differences. In the
presence of a strong phase difference, the rate asymmetry would
depend on both $\sin 2\beta$ and $\cos 2\beta$ and,
when combined with a $\sin 2\beta$ measurement from \BPSIKS\ decays,
could aid in the resolution of ambiguities in the determination of
$\beta$. 

 The decay \BDDST\ would also provide a clean test
of the factorization ansatz for decays into two charm mesons
and provide a measurement of the ratio of \DST\ and \DP\ decay 
constants and form factors~\cite{ref:pham,ref:xing.plb443}.

 The expected branching fractions of the decays \BDD\ can be
estimated from the measurement of the corresponding Cabibbo-favored
processes \BDSD~\cite{ref:pdg} and the ratio of 
decay constants~\cite{ref:rosner,ref:decay.constants}. 
The estimated \BDSTDST\ 
branching fraction is $\sim 10\TTMF$, consistent with
the measurement of 
$(6.2{}^{+4.0}_{-2.9}\ {\rm [stat.]}\pm 1.0\ {\rm [syst.]})\TTMF$~\cite{ref:first.obs},
and the estimates for \BDDST\ and \BDPDM\ are $\sim 8\TTMF$ and 
$\sim 5\TTMF$, respectively. 

 We present an update of the previous CLEO 
measurement of \BR(\BDSTDST)~\cite{ref:first.obs} and improved
upper limits on \BR(\BDDST) and \BR(\BDPDM)~\cite{ref:cleo.ul}
based upon a sample of $9.7\times 10^6$ \BB\ pairs produced in
\EPEM$\to$\UPSFOUR\ decays accumulated with the
CLEO detector at the Cornell Electron Storage Ring (CESR).
We also present the first angular analysis of \BDSTDST\ decays and
limit the $CP$-odd content of this reaction. The results
presented here supersede the previous CLEO results~\cite{ref:first.obs,ref:cleo.ul}.

\section{The CLEO detector\label{sec:cleo.det}}

 The data were accumulated with two configurations of the CLEO
detector dubbed \cleoii~\cite{ref:cleoii} and \cleoiiv~\cite{ref:cleoiiv}.
In the first configuration, a 1.5T solenoidal magnetic field encloses
three concentric cylindrical drift
chambers that are nested within a cylindrical barrel of time-of-flight (TOF)
scintillators and a CsI(Tl) calorimeter.
The surrounding iron return yoke is instrumented with proportional
wire chambers for muon identification. The large outer drift chamber
provides up to 49 measurements of a charged
particle's specific ionization ($dE/dx$) for particle species identification.
In the \cleoiiv\ configuration, the innermost wire chamber was replaced 
by a three-layer, silicon vertex detector (SVX) capable of providing
precision position information in both $r\phi$ and $z$~\cite{ref:coord.sys}.
The gas in the large outer drift chamber was also changed from
argon-ethane to helium-propane, resulting in improved $dE/dx$
and momentum resolution~\cite{ref:peterson}.

The Monte Carlo simulation of the CLEO detector response was
based upon GEANT~\cite{ref:geant}. Simulated events for the \cleoii\ 
and \cleoiiv\ configurations were processed in the same manner as
the data.

\section{Charm meson reconstruction\label{sec:charm.reco}}

Observation of the relatively small rates
expected for \BDD\ decays 
requires an aggressive program of charm meson reconstruction.
The \DZ\ decay modes considered for reconstruction
are \KPI, \KPIPIZERO, \KPIPIPI, \KSHORTPIPI\ and \KSHORTPIPIPIZERO;
the \DP\ decay modes considered for reconstruction are
\KPIPI, \KSHORTPI, \KSHORTPIPIZERO, \KSHORTPIPIPI, \KPIPIPIZERO,
\KKPI\ and \KKPIPIZERO. 
In order to limit background, 
the decay $\DP\to\KPIPIPIZERO$ is not considered for the
reconstruction of the \BDSTDST\ mode and the
decays $\DP\to\KKPI$ and $\DP\to\KKPIPIZERO$ are not considered for the
reconstruction of the \BDPDM\ mode.
The \DST\ decays to $\DZ\pi^+$ and $\DP\pi^0$ are selected for
the reconstruction of the \BDSTDST\ and \BDDST\ modes, 
 although
the final state $(D^+\pi^0)(D^-\pi^0)$ is overwhelmed by
combinatorial background and is excluded from 
the \BDSTDST\ reconstruction.
In the following, ``$D$'' refers to either \DP\ or \DZ\ mesons,
``\SLOWPI'' refers to the slow pion daughter of the \DST\ decay
and charge conjugation is implied unless explicitly stated otherwise.

 Charged kaon and pion daughters of $D$ meson candidates must be 
compatible with an origin at the \EPEM\ interaction point.
The $dE/dx$ or TOF measurement of a charged track, when available,
must be 
within 2.5 and 3.0 standard
deviations ($\sigma$) 
of expectations 
for $K^\pm$ and $\pi^\pm$ candidates, respectively.
The \KS\ meson candidates are reconstructed in the $\pi^-\pi^+$ decay
mode and must be consistent with an origin at the \EPEM\ interaction
point. At least one of the \KS\ daughter pions must be inconsistent with an
origin at the \EPEM\ interaction point. 
Neutral pion candidates are 
formed from energy deposits in the calorimeter consistent with
electromagnetic showers unassociated with a charged track
and with an energy exceeding
30 MeV in the barrel ($|\cos\theta|<0.71$) and 50 MeV in the endcap region
where $\theta$ is the angle of the shower with respect to the $z$ axis.
A requirement on the 
$\pi^0$ minimumn momentum of 100 MeV/$c$ is imposed for $D$ daughter candidates
and of 70 MeV/$c$ for \DST\ daughter candidates.
The charged and \KS\ daughters
of all $D$ meson candidates are required to originate from 
a common vertex.

\section{\BZ\ meson candidate selection\label{sec:b.reco}}

 A number of observables are used to suppress backgrounds.
In general, the requirements on these are more stringent
for the \BDDST\ and \BDPDM\ modes than for \BDSTDST\ 
because the combinatorial backgrounds are larger.
In addition, while common selection criteria for all
\DST\ and $D$ decay modes of each \BZ\ candidate
were
satisfactory for the \BDSTDST\ mode, the \BDDST\ and
\BDPDM\ modes require separate criteria for each \BOR\
channel~\footnote{For example, for the \BDPDM\ mode,
there are a total of 30 possible channels for the five \DP\ decay modes
in each detector configuration.} to reduce background.
The selection criteria for each channel of the 
\BDDST\ and \BDPDM\ modes were optimized using simulated signal and
background events assuming $\Br(\BDDST) = 8\TTMF$ and 
$\Br(\BDPDM) = 4.5\TTMF$, respectively.

\subsection{\BZ\ meson candidate energy and mass\label{subsec:mb.de}}

 The observable $\DELE\equiv E(D^{(*)+}) + E(D^{(*)-}) - E_{\rm beam}$
exploits energy conservation for \BZ\BZBAR\ meson pairs produced
in \UFOURS\ decays and 
has a resolution $\sigma(\DELE) = 8\ {\rm MeV}$ after constraining
the \BZ\ daughter candidates to the $D^{(*)+}$ masses~\cite{ref:pdg}.
The beam-constrained $B$ mass is defined as
$\MB^2 \equiv E^2_{\rm beam} - {\mathversion{bold}{\bf p}}_B^2$,
where ${\mathversion{bold}{\bf p}}_B$ is the measured \BZ\ candidate
momentum.
The \MB\ resolution of 2.5 MeV  is dominated by the beam energy
spread~\cite{ref:mb.resolution}. 
Signal candidates
are selected by requiring both \DELE\ and $\MB - M^{\rm n}_B$ to be within
$2.5\sigma$ of zero for the \BDSTDST\ mode and
within $2.0\sigma$ of zero for the \BDDST\ and \BDPDM\ modes, 
where $M^{\rm n}_B$ is the world-average \BZ\ mass~\cite{ref:pdg}.

\subsection{Candidate mass $\chi^2$ \label{subsec:masschi}}

The overall deviation  of \DST\ and $D$  candidates
from the \DST\ and $D$ meson masses is quantified by 

\begin{equation}\label{eqn:chim}
\CHIM \equiv \sum_i
\left(\frac{M_i - M_i^{\rm n}}{\sigma(M_i)}\right)^2
+
\left(\frac{\Delta M_i - \Delta M_i^{\rm n}}{\sigma(\Delta M_i)}\right)^2
\ \ \ ,
\end{equation}
\noindent where $M_i$ is the measured $D$ candidate mass,
$\Delta M_i$ is the mass difference between the \DST\ and
$D$ candidates, and
$\sigma(M_i)$ and $\sigma(\Delta M_i)$ are the corresponding 
resolutions. The superscript ``n'' denotes the world-average 
mass or mass difference~\cite{ref:pdg}. 
The sum runs over $i=D^{(*)+},D^{(*)-}$; 
the second term in Eqn.~(\ref{eqn:chim}) is not present for
\BDPDM\ candidates and is only present for the $i=\DSTP$ term for \BDDST\ 
candidates.
For \BDSTDST\ candidates, the
average resolutions were used; for \BDDST\ and \BDPDM\ 
decays, the resolution for each $D$ and \DST\ candidate was determined
from the track covariance matrices. If more than one \BDD\ 
candidate was present in a single event after all other selection
criteria were applied, the one with the smallest \CHIM\ was
selected. This observable is most effective for \BDSTDST\ since
$\sigma(\Delta M) \approx 500 \ {\rm keV}$ and $350\ {\rm keV}$
 for the $\DZ\pi^+$  final state in \cleoii\ and  \cleoiiv,
respectively. 
We require $\CHIM < 10 $ for \BDSTDST\ candidates~\cite{ref:first.obs}.
A typical requirement on \CHIM\ is $<\! 6$ for \BDDST (\DDSTtyp)
and $<\! 4$ for \BDPDM (\DPDMtyp).

\subsection{Separation between the $D$ and $\bar{D}$ decay vertices\label{subsec:lsl}}

The observable \LSL\ exploits the relatively
long decay length of the \DP\ meson ($\gamma\beta c\tau \approx 250\ \mu{\rm m}$)
and is defined as

\begin{equation}\label{eqn:lsl}
L
\equiv
( {\mathversion{bold}{\bf v}}_D - {\mathversion{bold}{\bf v}}_{\bar{D}} )
\cdot
\frac {( {\mathversion{bold}{\bf p}}_D - {\mathversion{bold}{\bf p}}_{\bar{D}})}
      {| {\mathversion{bold}{\bf p}}_D - {\mathversion{bold}{\bf p}}_{\bar{D}}|}
\ \  ,
\end{equation}
\noindent where ${\mathversion{bold}{\bf v}}_D$ 
(${\mathversion{bold}{\bf p}}_D$)
is the reconstructed $D$ candidate decay vertex (momentum).
The resolution $\sigma(L)$ is determined from the $D$ candidates'
covariance matrices; typically, $\sigma(L) = 500\ (200)\ \mu{\rm m}$
for \cleoii\ (\cleoiiv). For \cleoii, only the 2-dimensional
$r\phi$ information is precise enough to provide some discrimination
so we use only the $r\phi$ projection of $L$;
in \cleoiiv, the SVX allows the use of the full 3-dimensional
vertex information.
We require $\LSL > 0$ for $\BDSTDST\to(\DP\pi^0)(\bar{D}^0\pi^-)$ candidates in 
\cleoiiv\ only~\cite{ref:first.obs}.
For the \cleoiiv\ detector configuration, 
typical requirements on \LSL\ are $> -0.5$ for \BDDST (\DDSTtyp) 
and $\LSL > 2.5$ for \BDPDM (\DPDMtyp).

\subsection{Thrust and helicity angle\label{subsec:angle}}

For the \BDDST\ and \BDPDM\ modes,
the observable \COSTHR\ was used to suppress non-\BB\ background.
The angle between the thrust axis~\cite{ref:thrust} 
of the \BZ\ candidate and the thrust axis of the remainder of
the event is \THRA. Continuum ($\EPEM\to\QQ$, $q=u,c,s,d$) backgrounds are
sharply peaked towards $|\COSTHR| = 1$ and signal events are uniform
in \COSTHR. The maximum allowed $|\COSTHR|$ ranges from 0.50 to 0.95 for 
the \BDDST\ channels and from 0.80 to 0.95 for the  \BDPDM\ channels.

 The ${\rm pseudoscalar}\to {\rm vector}, {\rm pseudoscalar}$
decay \BDDST\ produces a $\cos^2\HELA$ distribution for signal
and is uniform for background. The angle \HELA\ is taken between the
\SLOWPI\ and the \DST\ in the \DST\ rest frame. 
The minimum allowed $|\COSHEL|$ for \BDDST\ candidates 
lies in the range $0.1$ to $0.7$, depending on the
decay channel.

\subsection{\DP\ decay length}

The \BDPDM\ mode suffers from a
background that consists of a 
\DP\ candidate where the majority
of daughter candidate tracks are the result of a \DP\ meson decay and
a \DM\ candidate composed of a random combination of tracks.
The observable \LSL\ (Sec.~\ref{subsec:lsl}) does not sufficiently
suppress this background due to the decay length of the \DP\ 
candidate,
but a requirement on 
$S\equiv \min(d_D/\sigma(d_D),d_{\bar{D}}/\sigma(d_{\bar{D}}))$,
the minimum decay length significance of the \BZ\ daughters,
where 
$d_D\equiv ( {\mathversion{bold}{\bf v}}_D - {\mathversion{bold}{\bf v}}_{B} )
         \cdot {\mathversion{bold}{\bf p}}_D /| {\mathversion{bold}{\bf p}}_D |$,
reduces this background component.
The average \BZ\ decay
length is $\sim 30\ \mu{\rm m}$; therefore, the \BZ\ decay vertex 
${\mathversion{bold}{\bf v}}_B$ can be accurately approximated as
the \EPEM\ interaction point. For the \KPIPI,\ \CKPIPI\ final state,
we require $S>-0.5$  for the \cleoiiv\ configuration.

 For the \BDDST\ and \BDPDM\ modes, there are channels
for which the background could not be reduced to a reasonable level
with any combination of selection criteria.
Specific \BDDST\ channels were discarded if the background estimated from
simulation could not be reduced
below $1/6$ of the expected signal rate.
Out of 84 possible channels considered,
 a total of 57 and 67 \BDDST\ channels
survive this criterion for the \cleoii\ and \cleoiiv\ detector
configurations, respectively.
Similarly, \BDPDM\ channels for which  the background
could not be reduced below $1/3$ or $1/7$ of the expected
signal rate for the \cleoii\ or \cleoiiv\ configuration, respectively,
were rejected. These criteria select 8 and 7 out of 
a total of 15 possible \BDPDM\ channels 
for the \cleoii\ and \cleoiiv\ configurations, respectively.

\section{Results and interpretation}

\begin{figure}
	\begin{center}
	\epsfig{file=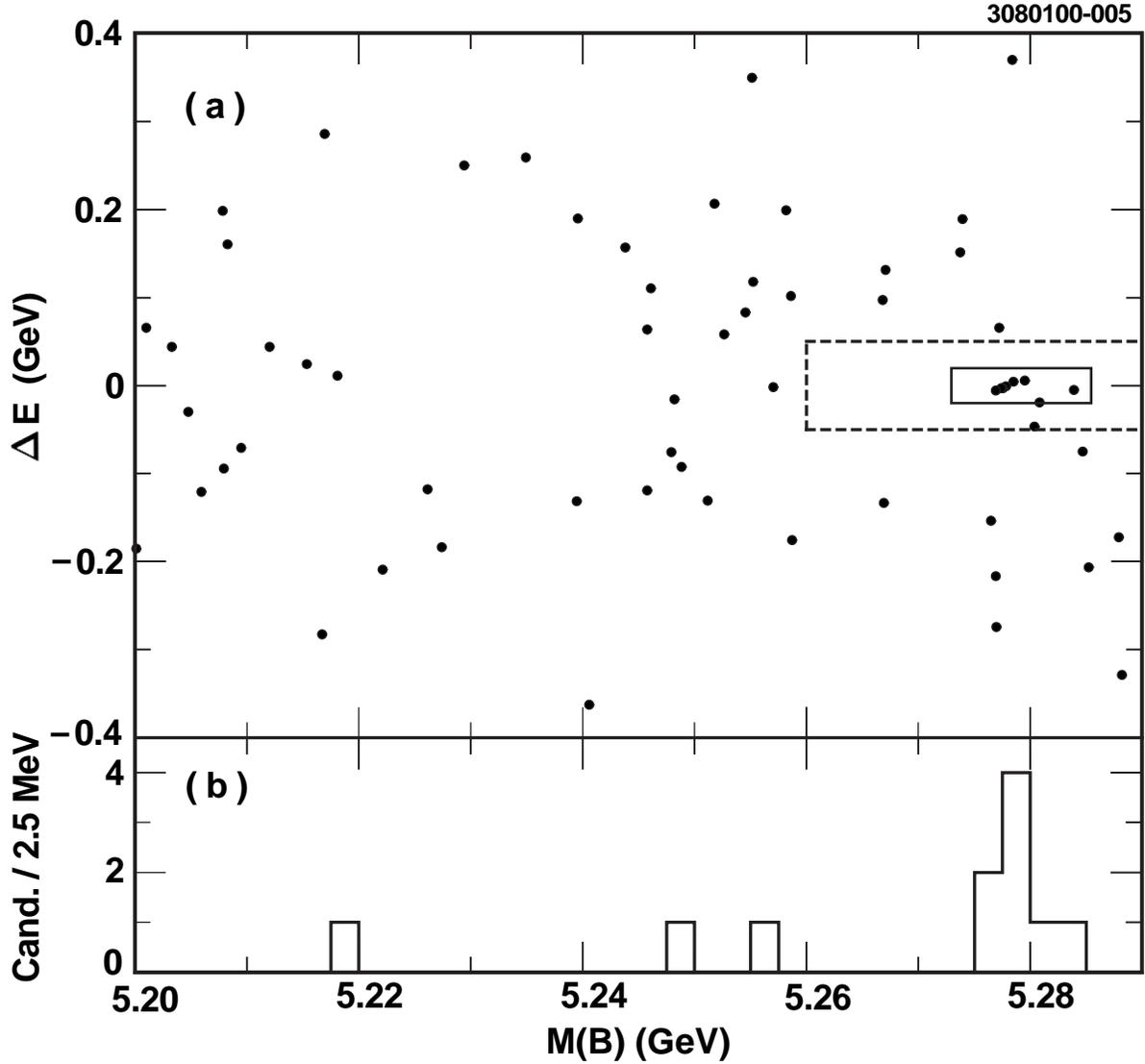,width=\linewidth}
	\caption{\label{fig:dstdst}
	(a) The \DELE\ {\it vs.} \MB\ distribution for \BDSTDST\ candidates
	for the data taken at the \UPSFOUR\ resonance.
	 The small rectangle
	delineates the signal region and the region outside the dashed line
	is the GSB.
	(b) The \MB\ distribution with the requirement $|\DELE| < 20 \ {\rm MeV}$.
	}
	\end{center}
\end{figure}

\begin{figure}
	\begin{center}
	\epsfig{file=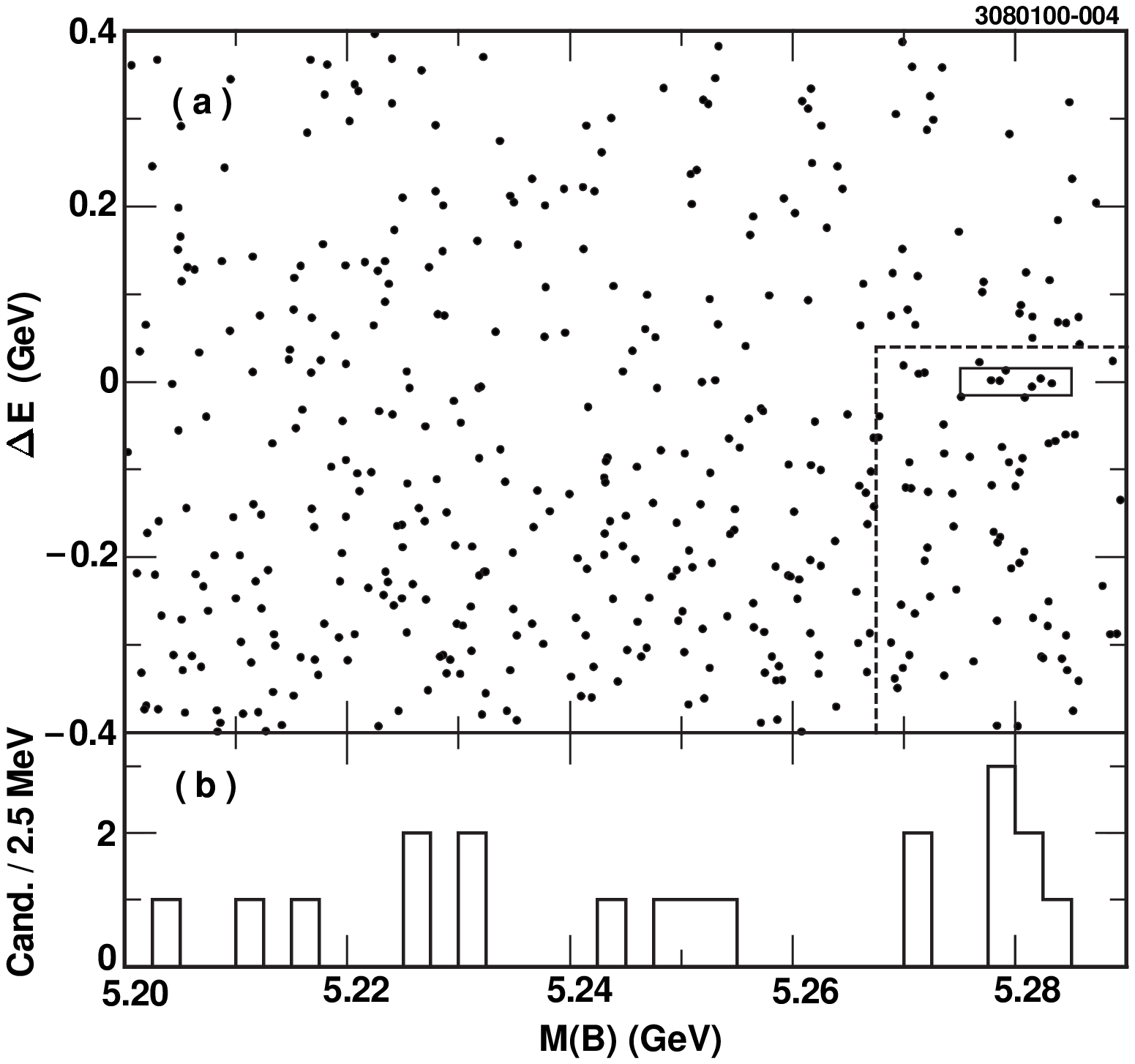,width=\linewidth}
	\caption{\label{fig:ddst}
	(a) The \DELE\ {\it vs.} \MB\ distribution for \BDDST\ candidates
	for the data taken at the \UPSFOUR\ resonance.
	 The small rectangle
	delineates the signal region and the region outside the dashed line
	is the GSB.
	(b) The \MB\ distribution with the requirement $|\DELE| < 16 \ {\rm MeV}$.
	}
	\end{center}
\end{figure}

\begin{figure}
	\begin{center}
	\epsfig{file=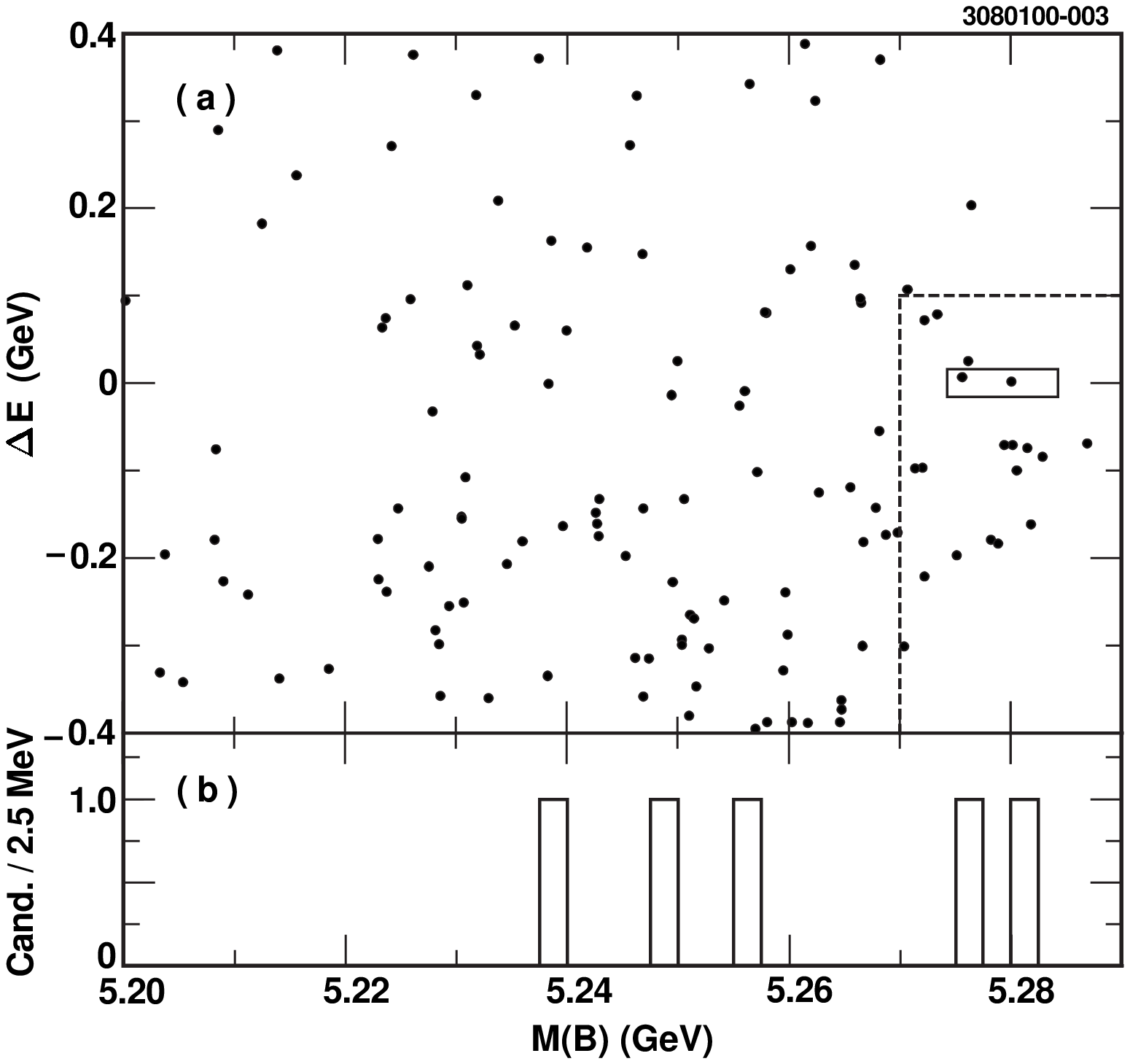,width=\linewidth}
	\caption{\label{fig:dpdm}
	(a) The \DELE\ {\it vs.} \MB\ distribution for \BDPDM\ candidates
	for the data taken at the \UPSFOUR\ resonance.
	 The small rectangle
	delineates the signal region and the region outside the dashed line
	is the GSB.
	(b) The \MB\ distribution with the requirement $|\DELE| < 16 \ {\rm MeV}$.
	}
	\end{center}
\end{figure}

 The \DELE\ {\it versus} \MB\ distributions of \BDSTDST, \BDDST\ and
\BDPDM\ candidates passing all selection criteria are shown in
Figures~\ref{fig:dstdst}, \ref{fig:ddst} and \ref{fig:dpdm},
respectively. A significant signal is apparent
for \BDSTDST\ decays; the larger backgrounds for the \BDDST\ and
\BDPDM\ modes are discussed below.

\subsection{Background estimation\label{subsec:bkgd}}

 For all three modes, the background is estimated with two
independent methods based on samples drawn largely from the 
data~\cite{ref:first.obs}.
Method 1 uses the grand sideband (GSB) indicated in 
Figures~\ref{fig:dstdst}, \ref{fig:ddst} and \ref{fig:dpdm}.
The observed number of candidates in the GSB in each channel is scaled
 to estimate the background in the signal region. The
scale factors are 
\SCFBDSTDST, \SCFBDDSTII, \SCFBDDSTIIV\ and \SCFBDPDM\ for the
 \BDSTDST, \BDDST(\cleoii), \BDDST(\cleoiiv) and
\BDPDM\ analyses, respectively, and are estimated from the fitted 
distributions in 
\MB\ and \DELE.
The excluded region of the GSB contains fully- or 
partially-reconstructed $B\to D^{(*)+}D^{(*)-}X$
decays that cannot enter the signal region.
 The GSB regions are 
slightly smaller for the \BOR\ analyses 
because they suffer from ``reflection'' background.
``Reflection'' backgrounds
arise if Cabibbo-favored \BDSD\ decays are interpreted as \BOR\ 
when  a charged kaon from the $D_{s}^{(*)+}$ is misidentified as a pion.
This background has $\DELE \le -50 \ {\rm MeV}$ due to the kinematics
of the $D_{s}^+$ decay combined with the difficulty in  distinguishing 
$K^\pm$ from $\pi^\pm$ for 
$|{\mathversion{bold}{\bf p}}| \ge 800\ {\rm MeV}/c$ with $dE/dx$ or TOF.

 For method 2 the
contribution of each background component was estimated
separately.
The dominant contribution to the background consists
of combinations of \DOPP\ and \DOPM\ in which one or both
candidates is fake; that is, the $D^{(*)}$ daughter candidates
are not the result of a $D^{(*)}$ meson decay. 
This combinatorial background can be estimated by
forming explicit fake \DOPP\ candidates 
drawn from the $D$ candidate mass sidebands
by replacing
$M_i^{\rm n}$ in Eqn.~\ref{eqn:chim} with 
$M_i^{\rm n} + f\sigma(M_i)$ or 
$M_i^{\rm n} - f\sigma(M_i)$.
We use $f=6$ so that classification of each 
$D$ meson candidate as fake or standard
is unique given the \CHIM\ selection criteria.
The contribution
to each channel of the combinatorial background can be derived
from the two samples consisting of fake \DOPP\ and standard \DOPM\ 
candidates or fake \DOPP\ and fake \DOPM\ candidates. 

Two  other background components are due to random
combinations of real \DOPP\ and \DOPM\ mesons that are approximately 
back-to-back and arise from the processes
$\EPEM\to\CC\to\DOPP\DOPM\ X$ or 
$\EPEM\to\UPSFOUR\to\BB\to (\DOPP X)(\DOPM Y)$.
The 
$\EPEM\to\CC\to\DOPP\DOPM\ X$ component was estimated from
$4.6 \ {\rm fb}^{-1}$ of \EPEM\ data taken 60 MeV below the
\UPSFOUR\ resonance after subtraction of the combinatorial background
using the method described above. 
The $\EPEM\to\UPSFOUR\to\BB\to (\DOPP X)(\DOPM Y)$ 
component was estimated from samples of simulated events at least
10 times the data sample size. 
The estimated total backgrounds
are listed in Table~\ref{tab:bkgd}. 
The estimates from the two methods for each 
channel are in good agreement and are combined 
channel-by-channel 
to produce
the overall background estimate.

\begin{table}
	\caption{\label{tab:bkgd}
		Background estimates. The two background
	estimation methods are described in the text.
	For method 2, the combinatorial, \CC\ and \BB\ components
	of the background are listed separately.
	The uncertainties in the table are statistical only and
 	do not include the uncertainty due to the 
	background scaling factor derived from the fitted 
	\DELE\ and \MB\ distributions (Sec.~\protect\ref{subsec:bkgd}).
		}
	\begin{tabular}{|l|c||c|ccc|}
         & Method 1     & \multicolumn{4}{c|}{Method 2}\\
Decay    & Total        & Total           & combinatorial & \CC\ & \BB\  \\
\hline 
\BDSTDST&$0.384\pm0.053$&$0.469\pm0.057$  & $0.382\pm0.046$ & $0.052\pm0.034$ & $0.035\pm0.005$  \\
\BDDST\ &$1.874\pm0.102$&$1.795\pm0.098$  & $1.336\pm0.062$ & $0.305\pm0.078$ & $0.064\pm0.005$  \\
\BDPDM\ &$0.498\pm0.048$& $0.459\pm0.041$ & $0.433\pm0.039$ & $0.014\pm0.003$ & $0.013\pm0.003$ \\
	\end{tabular}

\end{table}

 We assess the probability for the estimated background to
produce a more ``signal-like'' configuration of candidates
than the observed $B\to\DOPP\DM$ signal candidates with the likelihood
${\cal L} = \prod_i f(b_i;n_i)$, where the product runs
over all channels selected for either the \BDDST\ or
\BDPDM\ analysis, $f(\mu;n) \equiv e^{-\mu}\mu^n/n!$, 
$b_i $ is the estimated background in the $i^{\rm th}$
channel and $n_i  $ is the observed number of signal
candidates in the $i^{\rm th}$ channel. We compare the
distribution of ${\cal L}$ for many simulated experiments
consisting solely of background with the value of
${\cal L}$ obtained for the signal candidates in the data.
In the simulation of the background-only experiments, we
take into account both the statistical and systematic
uncertainty in the per-channel background estimates. 
For the \BDDST\ and \BDPDM\ mode, a total
of $0.3\%$ and $3.8\%$, respectively, of the simulated, background-only
experiments had ${\cal L} > {\cal L}_{\rm data}$ and, hence, are
more signal-like than the observed candidates. 
These rates are too large to claim an unambiguous observation
of either the \BDDST\ or \BDPDM mode. For the \BDSTDST\ mode,
fewer  than $2\times 10^{-7}$ background-only experiments were
more signal-like than the data.

\subsection{Branching fraction determination\label{subsec:bf}}

 The \BDD\ branching fractions are determined from the 
likelihood 
\begin{equation}\label{eqn:like}
{\cal L}(\BR) = \prod_i f(\mu_i;n_i)
\ \ ,
\end{equation}
\noindent where
\begin{itemize}
\item $\BR\equiv \BR(\BDD)$, 
\item $\mu_i = s_i + b_i$,
\item $s_i = 2 f_{00} N(\BB) \epsilon_i \BR_i(\DOPP) \BR(\BDD)$,
\item $\epsilon_i $ is the reconstruction efficiency of the 
      $i^{\rm th}$ channel, 
\item $\BR_i(\DOPP)$ is the product daughter branching fractions
of the $i^{\rm th}$ channel and
\item $N(\BB)$ is the number of \BB\ pairs.
\end{itemize}
We assume 
$f_{00}/f_{+-}\equiv \BR(\UPSFOUR\to\BZ\bar{B}^0)/\BR(\UPSFOUR\to B^+B^-) = 1$
for the results presented here. 
The evaluation of  ${\cal L}(\BR)$
takes into account the systematic uncertainties due
to the background estimate, efficiencies and \DOPP\ daughter
branching fractions~\cite{ref:pdg}. 
The branching fractions and
upper limits at 90\% CL for the three \BZ\ decay modes are listed in
Table~\ref{tab:results}. 
Since the background estimates of the two methods are combined
channel-by-channel, the combination of the total background estimates
of methods 1 and 2 (Table~\ref{tab:bkgd}) differs slightly from the
total background estimate given in Table~\ref{tab:results}.
Furthermore, the evaluation of the \BDD\ branching fractions
with a likelihood function that takes into account the
reconstruction efficiency, daughter branching fractions and backgrounds of
each channel (Eqn.~(\ref{eqn:like})) differs from the 
branching fraction that would be derived from 
the average efficiency times daughter branching fraction and total
backgrounds listed in Table~\ref{tab:results}.

  While only the \BDSTDST\ results provide
unambiguous evidence of the Cabibbo-suppressed
 $\bar{b}\to\bar{c}c\bar{d}$ decay, the
expectations based on the corresponding Cabibbo-favored decays
are consistent with the upper limits of the other two modes.
The results presented here indicate that there may be potential
 difficulties in the measurement of $\sin 2\beta$ using \BDD\ decays.
The yields are appreciably lower than that of \BPSIKS\ for the same 
integrated luminosity, and 
background levels are higher, especially for \BDDST\ and \BDPDM.
Measurement of $\sin 2\beta$ via the proper-time dependence of $\BZ\to\DOPP\DM$
decays performed at asymmetric \EPEM\ colliders or at hadron colliders
may be able to exploit the \BZ\ decay length to reduce backgrounds.
In contrast, the \BDSTDST\ results show that this mode,
while also having a yield substantially lower than that of 
\BPSIKS, has very low backgrounds and should provide an independent
measure of $\sin 2\beta$. 
The suppression of background for \BDSTDST\ is achieved largely through
the observable \CHIM (Sec.~\ref{subsec:masschi}) that relies on
accurate reconstruction of the trajectory of the charged slow pion
from the \DST\ decay. Inability to reconstruct efficiently the
$\pi_{\rm s}^+$ can substantially degrade a potential $\sin 2\beta$
measurement. For example, for the results presented here, the reconstruction
efficiency of the $\pi_{\rm s}^+$ from $\DST\to\DZ\pi_{\rm s}^+$ for
the \cleoiiv\ configuration is $(65\pm 6)\%$ of that for the \cleoii\ 
configuration because the track-finding algorithm was optimized only
for the latter configuration~\cite{ref:first.obs}.

\begin{table}[htb]
	\caption{\label{tab:results}
		The number of observed candidates, 
		estimated total backgrounds,
		efficiencies,
		measured branching fractions
		and branching fraction 
		upper limits at 90\% CL
	 	for the three \BDD\ modes.
		For the branching fractions and background, the first error
		is the statistical uncertainty and the second is the systematic
		uncertainty.
		$\langle\epsilon\BR\rangle$ is the product 
		of the reconstruction efficiencies and the $D^{(*)}$ daughter 
		branching fractions summed over all channels;
		the uncertainty includes both the statistical uncertainty in the
		estimation of $\epsilon_i$ from simulation as well as the
		uncertainties in the daughter branching fractions~\protect\cite{ref:pdg}.
		}
	\begin{tabular}{lccccc}
Decay     &          & Total          & $\langle\epsilon\BR\rangle$ & Branching        & 90\%CL Upper    \\
mode      &Candidates&  background    &  (\TTMF)                    & fraction (\TTMF) & limit (\TTMF)   \\
\hline 
\BDSTDST\ &   8      &\BKGDBDSTDST\  & \EFFBDSTDST\                & \BRBDSTDST\      & \ULBDSTDST\     \\
\BDDST\   &   6      &\BKGDBDDST\    & \EFFBDDST\                  & \BRBDDST\        & \ULBDDST\        \\
\BDPDM\   &   2      &\BKGDBDPDM\    & \EFFBDPDM\                  & \BRBDPDM\        & \ULBDPDM\        \\
\end{tabular}
\end{table}

\subsection{\BDSTDST\ transversity analysis\label{subsec:tv}}

\begin{figure}
	\begin{center}
	\epsfig{file=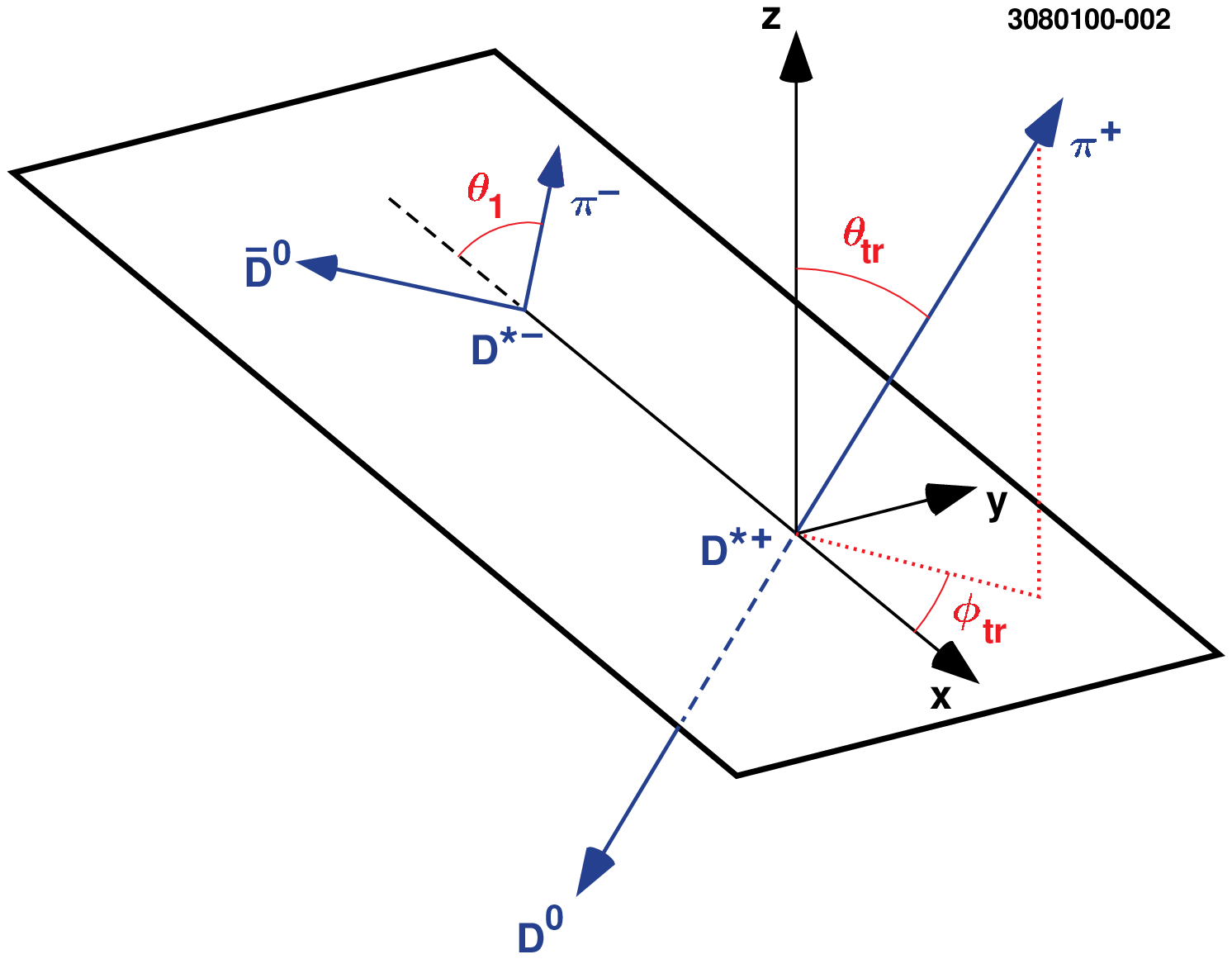,width=\linewidth}
	\caption{\label{fig:define.tv}
	The transversity frame for the decay \BDSTDST.
	}
	\end{center}
\end{figure}

A measurement of  $\sin 2\beta$ from \BDSTDST\ decays
requires an angular analysis to disentangle the $CP$-odd and $CP$-even
components of the decay. 
In the transversity basis~\cite{ref:dunietz}, the fraction 
of the $CP$-even component ($A$) of the decay \BDSTDST\ can
be determined from the \TV\ distribution, 

\begin{equation}\label{eqn:tv}
\frac {1}{\Gamma} \frac{d\Gamma}{d\TV} 
=
\frac{3}{4} A \sin^2\TVA
+
\frac{3}{2} (1-A)\cos^2\TVA
\ \ \ ,
\end{equation}
\noindent where $\Gamma \equiv \Gamma(\BDSTDST) + \Gamma(\BBARDSTDST)$
and \TVA\ is the angle between the \SLOWPI\ from the \DSTP\ and the
normal to the plane of the \DSTM\ decay in the \DSTP\ rest 
frame  as shown in Fig.~\ref{fig:define.tv}.

 We perform an unbinned, maximum likelihood fit to extract $A$
from the \TV\ distribution of the eight \BDSTDST\ candidates, taking
into account the background shape and the \TV\ resolution and
acceptance. The background shape, estimated from GSB candidates,
is consistent with being uniform as a function of \TV.
The resolution of $\sigma(\TV) = 0.1$ is determined from simulated
events in the observed decay channels. The acceptance varies 
as a function of \TV\ due to the drop in efficiency at low
momentum for the charged \SLOWPI. For $\pi^+_{\rm s}$ emitted
perpendicular (parallel) to the \DSTP\ direction, \TV\ tends towards
$\pm 1$ (0). Thus a loss of efficiency for low momentum 
$\pi^+_{\rm s}$ results in a reduction of acceptance at
\TV\ near zero. 
This effect is inconsequential for the $\DSTP\to\DP\pi^0$ candidates
because the $\pi^0_{\rm s}$ efficiency does not vary appreciably.
The acceptance is modeled as $\propto 1 + \alpha \cos^2\TVA$, where
$\alpha = 0.17\pm0.17$ is determined from simulated 
\BDSTDST\ decays and the  uncertainty represents a
conservative estimate of the range of $\alpha$.

\begin{figure}
  \begin{center}
   \epsfig{file=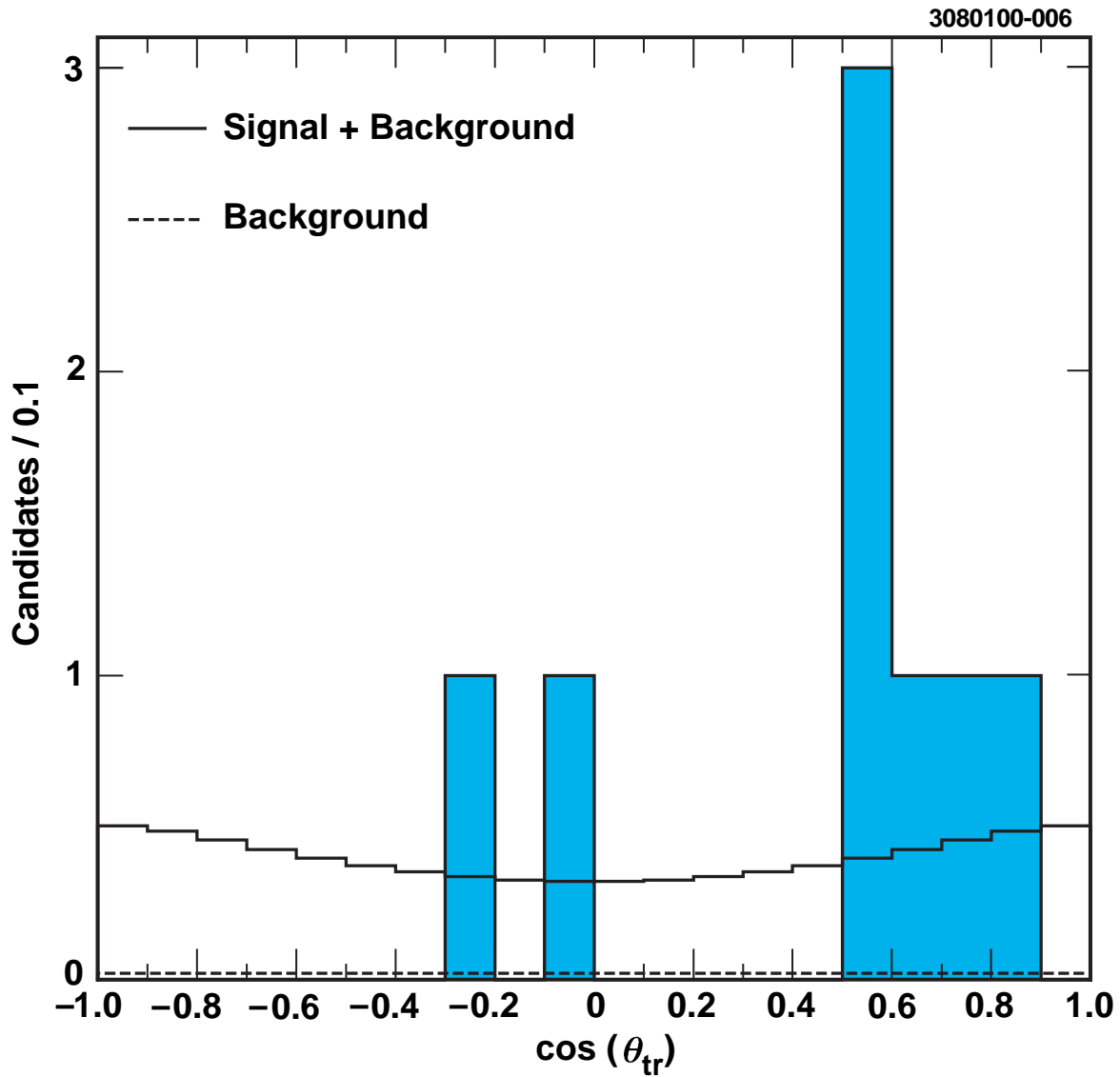,width=\linewidth}
   \caption{\label{fig:tv}
	The fitted  \TV\ distribution of the eight \BDSTDST\ 
        candidates from the signal region. The filled histogram represents the
	data, the solid line represents the best fit result and the 
	dashed line represents the background component.
	The fit takes into account the acceptance and resolution in \TV\ 
	as described in the text.
	      }
  \end{center}
\end{figure}

\begin{figure}[htbp]
  \begin{center}
   \epsfig{file=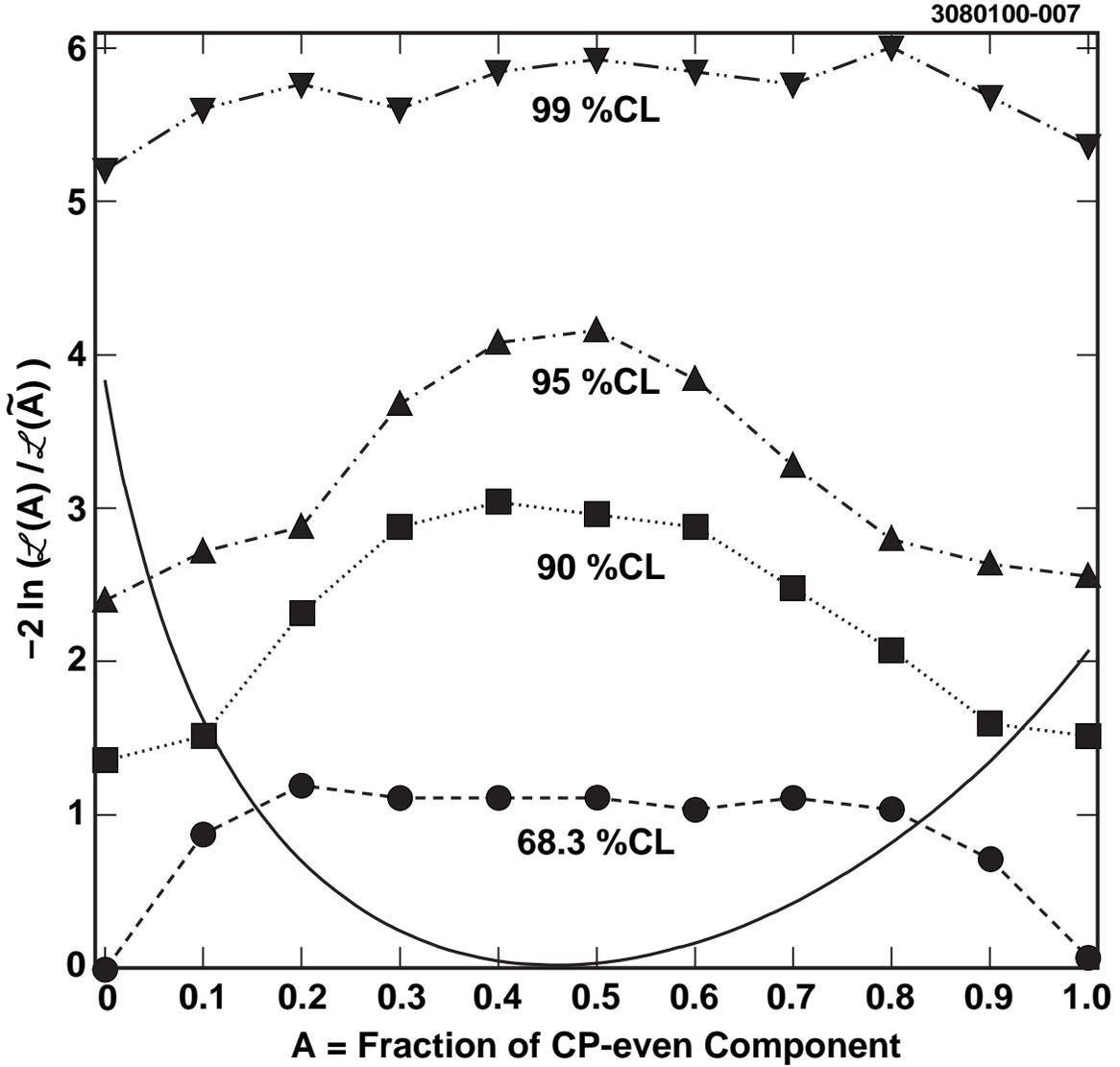,width=\linewidth}
   \caption{\label{fig:alike}
	$L_{\rm data}(A)  \equiv  -2\ln({\cal L}(A)/{\cal L}(\tilde{A}))$ for the data
	(solid curve) 
	compared to $L_n$ (Eqn.~(\protect\ref{eqn:L95})) for $n=$
	68.3\%, 90\%, 95\% and
	99\% (broken lines) that correspond to the confidence
	levels at $n$\% for $\alpha = 0.34$. See text for details.
	      }
  \end{center}
\end{figure}

 The observed \TV\ distribution of the \BDSTDST\ candidates
is shown in Fig.~\ref{fig:tv} with the fit result superimposed.
Figure~\ref{fig:alike} shows the dependence of 
$L(A) \equiv -2\ln({\cal L}(A)/{\cal L}(\tilde{A}))$, 
assuming $\alpha = 0.34$
where $\tilde{A}$ is the value of $A$ that maximizes ${\cal L}(A)$.
The conventional evaluation of confidence levels from
$L(A)$ is confounded because the statistical resolution 
on $A$ is comparable to the bounds on $A$ of $[0,1]$. To determine
confidence levels, we evaluate $L(A)$ as a function of the
input value of $A$ using 10000 simulated experiments at
each value of $A_{\rm input} = 0.0, 0.1, 0.2,\ldots,1.0$. Each
simulated experiment is analyzed as the data and the distribution
of $dN/dL(A_{\rm input})$ is determined ($N$ is the number of simulated
experiments). At each value of $A_{\rm input}$, we then 
determine the 95\% CL value,  $L_{95}$, as
\begin{equation}\label{eqn:L95}
{\int_0^{L_{95}} dL \, \frac{dN}{dL} }
\Big/ 
{\int_0^\infty  dL \, \frac{dN}{dL} }
= 0.95\ \ \ \ .
\end{equation}
In Fig.~\ref{fig:alike}
we show the curves resulting from this procedure at
the 68.3, 90, 95 and 99\% CL for $\alpha = 0.34$.
The  confidence level curves have a concave shape because
the $dN/dL$ distributions peak
more sharply at $A_{\rm input}$ near 0 and 1 due to the bounds
on $A$. 
We perform this procedure for the central and extreme
values of the acceptance, $\alpha = 0.00, 0.17, 0.34$, 
for both the simulation and the data to take into account
the acceptance uncertainty. We conservatively use the regions
excluded by all three values of $\alpha$ to set limits.
We exclude values of $A< 0.11$ at 90\% CL, but
cannot exclude $A=0$ at 99\% CL. Combining the limits
$0.15< A < 0.90$ at 68.3\% CL for the three values of $\alpha$ 
with the most likely value of $A$
for $\alpha=0.17$ 
and taking into account the uncertainties in the level and
shape of the background, we find $A = \ACP$.
Our results are consistent with expectations that 
$A\approx 0.95$\cite{ref:rosner,ref:explain,ref:pham}, although the statistical precision
is poor.

\section{Summary and conclusions\label{sec:conclusion}}

 We have studied the decays 
\BDSTDST, \BDDST\ and 
\BDPDM in
$9.7\times 10^6$ $\UPSFOUR\to\BB$ decays. 
We determine
$\BR(\BDSTDST) = (\BRBDSTDSTss)\TTMF$ and limit
$\BR(\BDDST) < \ULBDDST\TTMF$ and 
$\BR(\BDPDM) < \ULBDPDM\TTMF$ at 90\% CL.
These results, while consistent with expectations, show that 
substantially higher luminosities will be needed to make
a measurement of $\sin 2\beta$ using \BDD\ decays that
approaches the statistical precision of a $\sin 2\beta$
measurement using \BPSIKS. 
Asymmetry measurements of lesser precision with \BDD\ decays may,
however, be adequate for resolving ambiguities in the determination
of $\beta$.
We have performed the first transversity
analysis for \BDSTDST\ and exclude values of the $CP$-even
component of the decay less than \ANINETY\ at 90\% CL.

\section{Acknowledgments\label{sec:ack}}

We gratefully acknowledge the effort of the CESR staff in providing us with
excellent luminosity and running conditions.
I.P.J. Shipsey thanks the NYI program of the NSF, 
M. Selen thanks the PFF program of the NSF, 
M. Selen and H. Yamamoto thank the OJI program of DOE, 
M. Selen and V. Sharma 
thank the A.P. Sloan Foundation, 
M. Selen and V. Sharma thank the Research Corporation, 
F. Blanc thanks the Swiss National Science Foundation, 
and H. Schwarthoff and E. von Toerne
thank the Alexander von Humboldt Stiftung for support.  
This work was supported by the National Science Foundation, the
U.S. Department of Energy, and the Natural Sciences and Engineering Research 
Council of Canada.

\end{document}